\documentclass{PoS}
\rightline{\parbox{4cm}{\large\rm 
DESY 07-180\\
Edinburgh 2007/35\\
LTH 781
}}
\vspace{3.2cm}

\usepackage{graphicx}
\usepackage{amsmath}
\usepackage{txfonts}
\usepackage{amssymb}
\usepackage{amsfonts}
\usepackage{latexsym}

\newenvironment{Eqnarray}%
          {\arraycolsep 0.14em\begin{eqnarray}}{\end{eqnarray}}

\newcommand{\bc}{\begin{center}}
\newcommand{\ec}{\end{center}}
\newcommand{\eq}{\begin{equation}}
\newcommand{\ee}{\end{equation}}
\newcommand{\ea}{\begin{Eqnarray}}
\newcommand{\eea}{\end{Eqnarray}}

\newcommand{\emdh}{\mbox{$e^{-\Delta H}$}}
\newcommand{\bnfff}{\mbox{${\bf N_f}$=2+1}}
\newcommand{\nfff}{\mbox{$N_f$=2+1}}
\newcommand{\nff}{\mbox{$N_f$=2}}

\title{A status report of the QCDSF $\bnfff$ Project}
\ShortTitle{A status report of the QCDSF $\nfff$ Project}

\author{
        Meinulf G\"ockeler$^{a}$, Roger Horsley$^b$,
        \speaker{Yoshifumi Nakamura}$^c$, 
        Holger Perlt$^d$,
        Dirk Pleiter$^c$, Paul E.~L. Rakow$^e$,
        Gerrit Schierholz$^{cf}$, 
        Arwed Schiller$^d$,
        Thomas Streuer$^g$, Hinnerk St\"uben$^h$ and
        James M. Zanotti$^b$ \\
        \llap{$^a$} Institut f\"ur Theoretische Physik,
                    Universit\"at Regensburg, 
                    93040 Regensburg, Germany \\
        \llap{$^b$} School of Physics, University of Edinburgh, 
                    Edinburgh EH9 3JZ, UK \\
        \llap{$^c$} John von Neumann Institute NIC / DESY Zeuthen, 
                    15738 Zeuthen, Germany \\
        \llap{$^d$} Institut f\"ur Theoretische Physik,
                    Universit\"at Leipzig,
                    04109 Leipzig, Germany\\
        \llap{$^e$} Department of Mathematical Sciences,
                    University of Liverpool, 
                    Liverpool L69 3BX, UK \\
        \llap{$^f$} Deutsches Elektronen-Synchrotron DESY, 
                    22603 Hamburg, Germany \\
        \llap{$^g$} Department of Physics and Astronomy, University of Kentucky,
                    Lexington KY 40506, USA \\
        \llap{$^h$} Konrad-Zuse-Zentrum f\"ur Informationstechnik Berlin, 
                    14195 Berlin, Germany \\
        E-mail:  \email{yoshifumi.nakamura@desy.de}
   }

\author{QCDSF Collaboration}

\abstract{
We report about on-going simulations of $\nfff$ lattice QCD.
We use a tadpole improved Symanzik gauge action
and stout link smeared Wilson fermions with a clover term. We employ
the Hasenbusch trick for the degenerate u- and d-quarks, and the RHMC algorithm for the
simulation of the strange quark. 
}

\FullConference{XXVth International Symposium on Lattice Field Theory\\
                July 30- August 4, 2007\\
                University of Regensburg, Germany}
\begin{document}


\section{Introduction}

Over the past few years the QCDSF Collaboration has focused on
 simulations of lattice QCD with 2 flavors of dynamical quarks.
The real world consists, however, of $\nfff$ light quarks (up, down and strange).
We therefore extend our previous simulations to $\nfff$ 
where we continue to investigate
 hadron and quark masses, weak matrix element, hadron form factors,
moments of parton distributions as well as a variety of other key parameters
of the Standard Model. Our ultimate goal is to bring the systematic uncertainties
down to or below the experimental errors.

JLQCD found an unexpected first-order phase transition in the strong coupling regime
at relatively heavy quark masses when they employed the plaquette gauge action and the 
$O(a)$-improved Wilson fermion action in three-flavor QCD simulations~\cite{JLQCDnf31st}.
Using an improved gauge action should give us significantly better control on the 
continuum extrapolation.
Additionally, it is important to reduce somehow the chiral symmetry breaking arising from 
the Wilson fermion formulation. 
A well-known method to attenuate this symmetry breaking is adding
 a clover term.
The {\it UV filtering} method, which involves replacing covariant
 derivatives in the fermion action by smeared descendents, is also becoming standard. 
We employ a tadpole improved Symanzik gauge action and stout link smeared Wilson fermions 
 with a clover term. 
We also improve the algorithm to reduce simulation costs.

\section{The Action}

The tadpole-improved Symanzik action we use for $\nfff$ simulations is
\begin{equation}
S_G =
\frac{6}{g^2}\left[ 
   c_0 \sum_{\rm plaquette}\frac{1}{3}\, \mbox{Re}\,\mbox{Tr}\, (1-U_{\rm plaquette}) 
 + c_1 \sum_{\rm rectangle}\frac{1}{3}\, \mbox{Re}\,\mbox{Tr}\, (1-U_{\rm rectangle}) 
             \right] \,,
\end{equation}
where the coefficients $c_0$, $c_1$ are taken from tadpole improved
perturbation theory:
\begin{equation}
\frac{c_1}{c_0} = - \frac{1}{20 u_0^2} \,,
\end{equation}
with $c_0 + 8 c_1  = 1$, where
$u_0 = \left(\frac{1}{3} {\rm Tr}\, \langle U_{\rm plaquette} \rangle \right)^{\frac{1}{4}}$.
We write $\beta = \frac{6}{g^2} \, c_0$.
In the classical continuum limit $u_0 \rightarrow 1$ the coefficients 
assume the tree-level Symanzik values~\cite{Symanzik} $c_0=5/3$, $c_1=-1/12$.

We continue to use clover fermions with the action
\begin{equation}
\begin{split}
S_F = \sum_x \Big\{\bar{\psi}(x)\psi(x)
&- \kappa\, \bar{\psi}(x) U_\mu^\dagger(x-\hat{\mu})[1+\gamma_\mu]
\psi(x-\hat{\mu})\\[0.0em]
&- \kappa\, \bar{\psi}(x) U_\mu(x)[1-\gamma_\mu]
\psi(x+\hat{\mu})
+\frac{i}{2} \kappa \, c_{\rm SW}\, \bar{\psi}(x)
\sigma_{\mu\nu} F_{\mu\nu}(x) \psi(x) \Big\} \, ,
\end{split}
\end{equation}
but replace the gauge links $U_\mu$ 
in all terms of the fermion action except the clover term 
by stout
links~\cite{Morningstar}
\begin{equation}
U_\mu \rightarrow \tilde{U}_\mu(x) = e^{iQ_\mu(x)} \, U_\mu(x)\,,
\end{equation}
with
\begin{equation}
Q_\mu(x)=\frac{\alpha}{2i} \left[V_\mu(x) U_\mu^\dagger(x) -
  U_\mu(x)V_\mu^\dagger(x) -\frac{1}{3} {\rm Tr} \,\left(V_\mu(x)
  U_\mu^\dagger(x) -  U_\mu(x)V_\mu^\dagger(x)\right)\right] \,,
\end{equation}
where $V_\mu(x)$ is the sum over all staples associated with the link.
We take $\alpha = 0.1$ and perform 1 level of smearing, 
corresponding to a mild form of {\it UV filtering}~\cite{UVfilterDurr}.
In this status report we present results where we used
the tree-level value for the improvement coefficient, i.e.
$c_{\rm SW} = 1$, or used a value obtained from
tadpole-improved perturbation theory:
\begin{equation}
c_{\rm SW}=\frac{1}{u_{0}^{3}} [1 + g^2 (0.00706281 + 1.142004 \alpha - 4.194470 \alpha^2) ] \, ,
\hspace{3mm}
g^2=\frac{6}{\beta} \frac{20 u_0^2}{ 20 u_0^2 - 8 } \, .
\end{equation}
 Note that in the future we will use the results presented in~\cite{Holger07100990}.

This action has many advantages over our previously used one. 
In particular, due to {\it UV filtering}, it is expected to have
better chiral properties~\cite{hep-lat/0405026} and
 smaller cut-off effects~\cite{hep-lat/0405015}.
One may also hope that the tadpole-improved perturbative value of $c_{\rm SW}$
 is close to the non-perturbative value.


\section{The Algorithm}

The standard partition function for $\nfff$ improved Wilson fermions is 
\begin{equation}
\begin{split}
Z&=\int DU D\bar{\psi} D\psi e^{-S} \,,\\
S&=S_g(\beta) + S_{l}(\kappa_{l},c_{\rm SW}) + S_s(\kappa_{s},c_{\rm SW})\,,
\label{standard}
\end{split}
\end{equation}
where $S_g$ is a gluonic action, $S_l$ is an action for the degenerate u- and d- quarks
 and $S_s$ is an action for the strange quark. After integrating out fermions 
\begin{equation}
  S=S_g(\beta) - \ln  [\det M_{l}^{{\dagger}}M_{l}] [\det M_{s}^{{\dagger}}M_{s}]^{1\over 2}\,.
\end{equation}
We first apply even-odd preconditioning:
\begin{equation}
\det M_{l}^{\dagger} M_{l}
\propto \det(1+T_{oo}^{l})^2 \det Q_{l}^{\dagger}Q_{l} \,,
 \hspace{3mm}
~[\det M_{s}^{\dagger} M_{s}]^{1 \over 2}
\propto \det(1+T_{oo}^{s}) [\det Q_{s}^{\dagger}Q_{s}]^{1 \over 2} \,,
\end{equation}
where 
\begin{equation}
 Q = (1+T)_{\rm ee} - M_{\rm eo} (1+T)^{-1}_{\rm oo} M_{\rm oe} \, ,
 \hspace{3mm}
 T = \frac{\rm i}{2} c_{\rm SW}\, \kappa\, \sigma_{\mu\nu} F_{\mu\nu} \,.
\end{equation}
We then separate $\det Q_{l}^{\dagger}Q_{l} $ following Hasenbusch~\cite{Hasenbusch:2001ne}
\begin{equation}
\det Q_{l}^{\dagger}Q_{l}
= \det W_{l}^{\dagger} W_{l} 
\det {Q_{l}^{\dagger}Q_{l} \over W_{l} W_{l}^{\dagger} }\,,
~~~~~~~~~W=Q+\rho\,.
\end{equation}
Finally we modify the standard action to
\begin{equation}
S=S_g + S_{det}^{l} + S_{det}^{s} + S_{f1}^{l} + S_{f2}^{l} + S_{fr}^{s} \,,
\end{equation}
where
\begin{equation}
\begin{split}
&S_{det}^{l}=-2\, {\rm Tr}\, \log[1+T_{oo}(\kappa^l)]\, ,
 \hspace{3mm}
 S_{det}^{s}=- \, {\rm Tr}\, \log[1+T_{oo}(\kappa^s)]\, , \\
&S_{f1}^{l}=\phi_1^{\dagger}           [W(\kappa^l)^{\dagger}W(\kappa^l)]^{-1}  \phi_1 \, ,
 \hspace{3mm}
 S_{f2}^{l}=\phi_2^{\dagger}W(\kappa^l)[Q(\kappa^l)^{\dagger}Q(\kappa^l)]^{-1}
                                        W(\kappa^l)^{\dagger}\phi_2 \, , \\
&S_{fr}^{s}=\sum_{i=1}^n \phi_{2+i}^{\dagger}
           [Q(\kappa^s)^{\dagger}Q(\kappa^s)]^{-{1\over 2n}} \phi_{2+i} \,.
\end{split}
\label{split1}
\end{equation}
We calculate $S_{fr}$ 
using the RHMC 
algorithm~\cite{RHMC} with optimized values for $n$ and the number of fractions.
We now split each term of the action into one ultraviolet
and two infrared parts,
\begin{equation}
S_{\rm UV} = S_{g}\,, \hspace{3mm}
S_{\rm IR-1} = S_{det}^{l} + S_{det}^{s} + S_{f1}^{l}\,, \hspace{3mm}
S_{\rm IR-2} = S_{f2}^{l} + S_{fr}^{s}\,.
\end{equation}
In~\cite{AliKhan:2003br} we have introduced two different time
scales~\cite{Sexton:1992nu} for the ultraviolet and infrared parts of the
action in the leap-frog integrator. Here we shall go a step further and put
$S_{\rm UV}$, $S_{\rm IR-1}$ and $S_{\rm IR-2}$ on {\it three separate} time
scales,
\begin{equation}
\begin{split}
&V(\tau) =
\Big[ V_{\rm IR-2} \left({\delta\tau \over 2}\right)
      ~~A^{m_1}~~ 
      V_{\rm IR-2} \left({\delta\tau \over 2}\right)\Big] ^{n_\tau} \,,\\
&A =V_{\rm IR-1}\left({\delta\tau \over 2m_1}\right)
      ~~B^{m_2}~~ 
    V_{\rm IR-1}\left({\delta\tau \over 2m_1}\right) \,,\\
&B =V_{\rm UV}\left({\delta\tau \over 2m_1 m_2 }\right)
    V_Q       \left({\delta\tau \over  m_1 m_2 }\right)
    V_{\rm UV}\left({\delta\tau \over 2m_1 m_2 }\right)\,,
\end{split}
\end{equation}
where n$_\tau =\tau / (\delta \tau)$ and the $V$s are evolution operators of the Hamiltonian.
The length of the trajectory $\tau$ is taken to be equal to one in our simulations.

\section{Test calculations}

\begin{figure}[!thb]
\vspace{-2mm}
\includegraphics[angle=270,scale=0.28,clip=true]{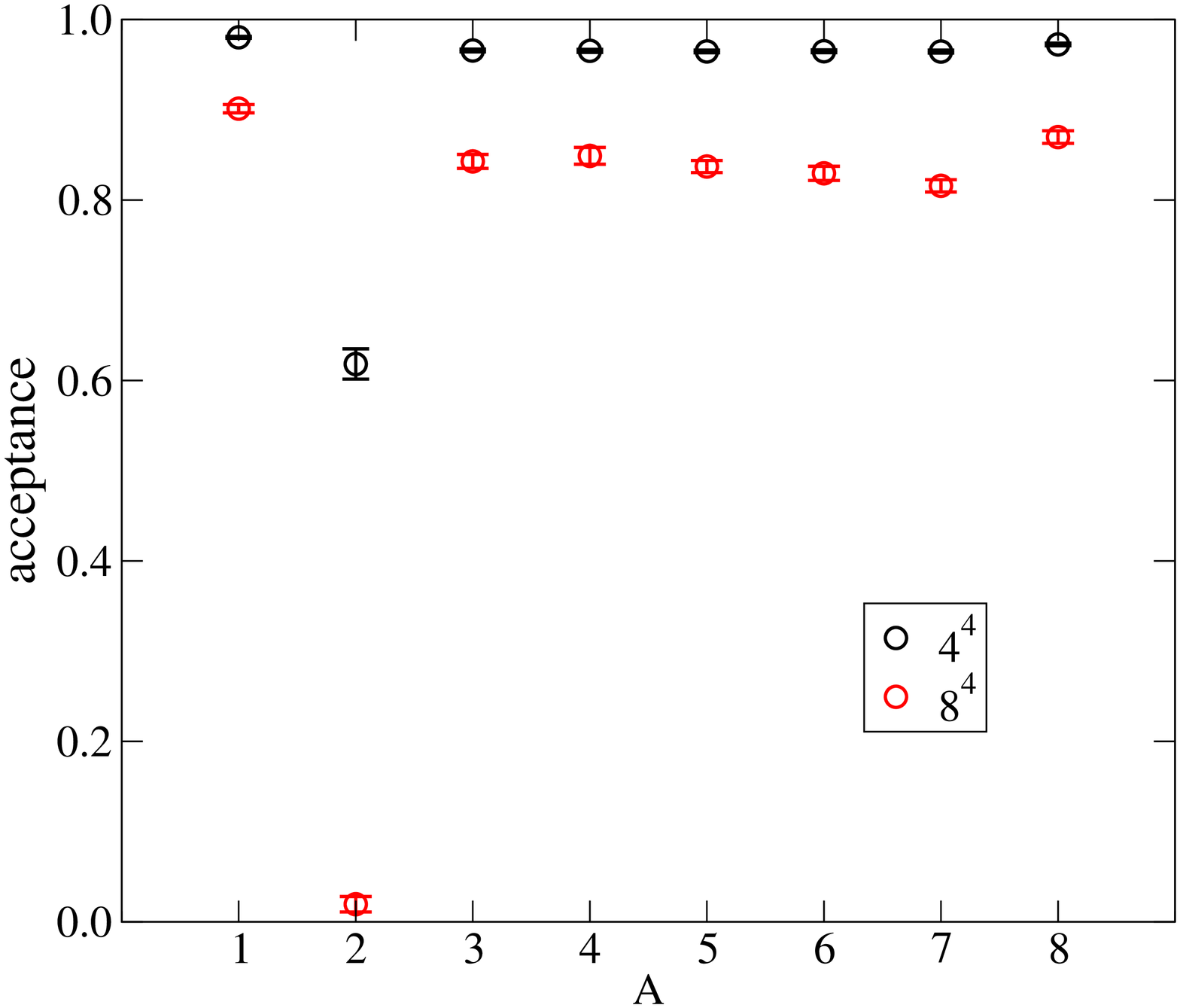}
\includegraphics[angle=270,scale=0.28,clip=true]{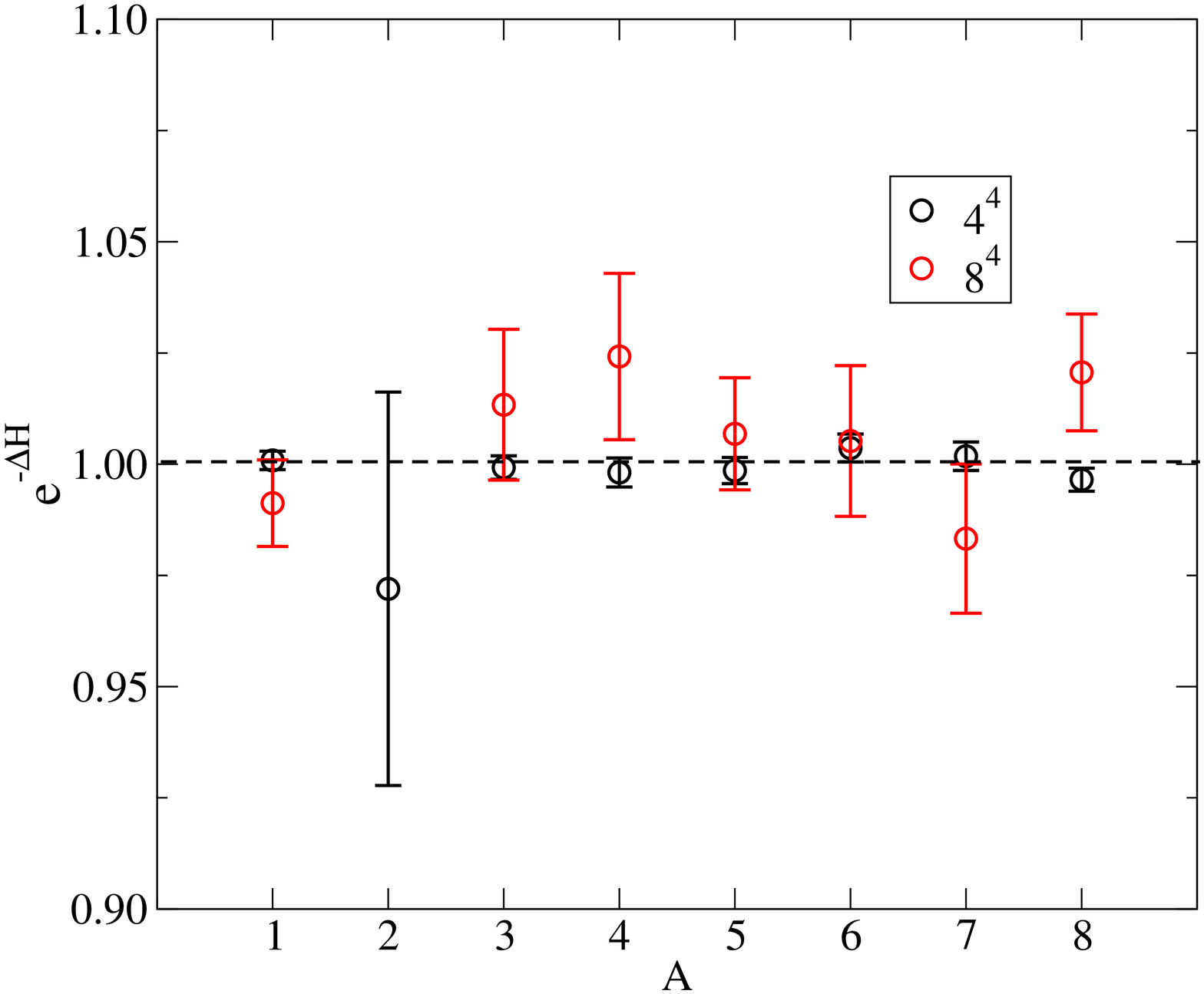}
\vspace{-3mm}
\caption{
The acceptance (left) and $\emdh$(right) for various simulation parameters.
A:($m_1$,$m_2$,$\rho$)=
(3,3,0.1),
(1,1,0.1),
(2,2,0.1),
(2,2,0.2),
(2,2,0.3),
(2,2,0.4),
(2,2,0.5),
(3,3,0.5).
The other simulation parameters are {f}{i}{x}{e}{d} to $n_{\tau}$=20, 
CG residual for Monte Carlo 
res$_{mc}$=$10^{-10}$, 
CG residual for Molecular Dynamics res$_{md}$=$10^{-7}$
 and the rational approximation by $n$=2, 20 fractions and range [0.01,3].
}
\label{acce2mdh}
\end{figure}

\begin{figure}[!thb]
\vspace{-3mm}
\bc
\includegraphics[angle=270,scale=0.3,clip=true]{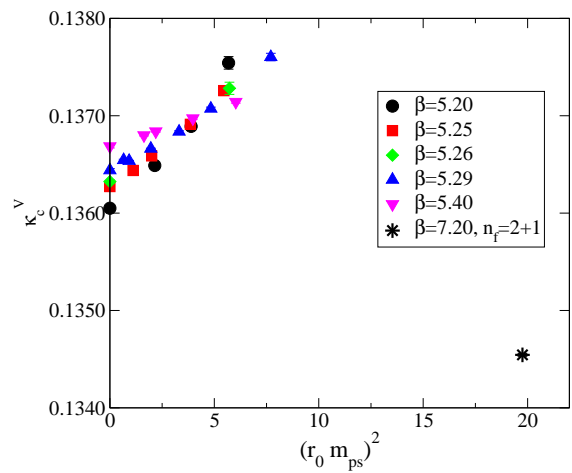}
\ec
\vspace{-5mm}
\caption{
$\kappa_c^V$ versus $(r_0 m_{PS})^2$ at $\beta=5.2 \sim 5.4$ for $\nff$ 
and at $\beta=7.2$ for $\nfff$.
}
\label{r0mps_kvc}
\end{figure}

We first tested our algorithm on small lattices of size $4^4$ and $8^4$.
Figure \ref{acce2mdh} shows the acceptance ratio and $\emdh$ for $\beta=7.2$, $\kappa_l=\kappa_s=0.1245$
and $c_{\rm SW}=1.0$ for various simulation parameters. 
We discarded for themalisation the first 200 trajectories and then 
calculated for each choice of the simulation parameters about 
1000 trajectories. 
 $\emdh$ should be equal to one within error and this is a good 
indicator of 
the correctness of the program.
As seen in Fig.~\ref{acce2mdh}, we can keep high acceptance and 
$\emdh$$\approx$1 by tuning parameters.
Simulation results for the same parameters on a $16^3~32$ lattice are presented 
in table \ref{tabk1245}. 
The lattice spacing roughly corresponds to the 
so-called {\it fine} run of the MILC collaboration~\cite{MICLfine} 
which is using the same gauge action.
Figure \ref{r0mps_kvc} is a plot of $\kappa_c^V$, which is obtained by extrapolating
the valence pion mass to zero,
 as a function of $(r_0 m_{PS})^2$. We see that $\kappa_c^V$ for $\nfff$ is lower than
 for $\nff$. 
Assuming that $\kappa_c^V$ also has a rather mild $\beta$
dependence, we may consider this as an indication that our
new action is much more continuum like.
\begin{table}[!thb]
  \begin{center}
  \begin{tabular}{ccccc}
    \hline\hline%
 $r_0/a$  & $am_{PS}$  & $am_V$     &$am_N$     & $\kappa_c^V$  \\ \hline
 4.89(15) & 0.9089(20) & 0.9520(25) &1.4828(53) & 0.134599(63)  \\
    \hline\hline
  \end{tabular}
  \end{center}
  \caption{
$r_0/a$, $m_{PS}$, $m_V$, $m_N$ and $\kappa_c^V$ obtained from partially quenched calculations 
on a $16^3~32$ lattice for $\beta=7.2$, $\kappa_l=\kappa_s=0.1245$ and $c_{\rm SW}=1.0$.
}
\label{tabk1245}
\end{table}

Results for $\beta=7.2$, $\kappa_l=\kappa_s=0.1335$ with tadpole improved $c_{\rm SW}$ on a $16^3~32$ lattice
are presented in table \ref{tabk1335}. 
They were calculated from 200 trajectories varying $n$ of eq. (\ref{split1}),
 the number of fractions and the precision of the coefficients 
for the rational approximation. 
Our results confirm that double precision coefficients are needed 
to obtain the correct value of $\emdh$. Note that 
the average plaquette value as well as the average minimum 
and maximum eigenvalue are consistent for the different 
choices of the algorithmic parameters.
But since $S_{fr}$ may be $O(10^8)$ (e.g. on large
lattices) the algorithm may not be correct when using
single precision coefficients.
For the parameters shown in the last row of table \ref{tabk1335}, 
the ratios for the force contributions from the different terms in the action are
\begin{equation}
{F_{det_l} \over F_{det_s}}=2       \,,\hspace{3mm}
{F_{f1}    \over F_{det_s}} \sim 30 \,,\hspace{3mm}
{F_{f2}    \over F_{det_s}} \sim 10 \,,\hspace{3mm}
{F_{fr}    \over F_{det_s}} \sim 10 \,,\hspace{3mm}
{F_{g}     \over F_{det_s}} \sim 90 \,.
\end{equation}
$F_{fr}$ for $n$=4 is 40$\%$
smaller than for $n$=2.

\begin{table}[!thb]
  \begin{center}
  \begin{tabular}{ccccccccc}
    \hline\hline%
$n$ & fr. & pr. &
$P$             & $\tau_{int}$    &
$\emdh$         & $P_{acc}$       &
$\lambda_{min}$ & $\lambda_{max}$ \\ \hline
2&32&s& 0.625591(55)& 2.07(76)& 1.167(25)& 0.96& 0.011247(64)& 2.4788(44)   \\
4&40&s& 0.625578(52)& 2.10(85)& 1.529(24)& 0.99& 0.011306(62)& 2.4757(40)   \\
2&32&d& 0.625528(41)& 1.74(46)& 0.995(17)& 0.91& 0.011316(66)& 2.4805(33)   \\
    \hline\hline
  \end{tabular}
  \end{center}
  \caption{
Simulation parameters and results for the value of the plaquette, the 
integrated autocorrelation time of the plaquette,
 $\emdh$, acceptance, minimum and maximum eigenvalues of $Q^{\dagger}Q$
on a $16^3~32$ lattice for $\beta=7.2$, $\kappa_l=\kappa_s=0.1335$ and
 tadpole improved $c_{\rm SW}$=1.612.
The parameters are the number $n$ of eq. (3.7),
 the number of fractions (fr.) and the precision of 
the coefficients for the rational approximation (pr.). The other parameters are {f}{i}{x}{e}{d} to $n_\tau$=60,
$m_1$=3, $m_2$=3, res$_{mc}$=$10^{-10}$, res$_{md}$=$10^{-8}$ and $\rho$=0.1.
}
\label{tabk1335}
\end{table}

\section{Conclusion}

In this contribution we have presented the status of our $\nfff$ project.
We found indications for our action to be better than our 
previously used action. Furthermore, we tested the correctness 
of our algorithm.
The performance of our program for matrix multiplication is about 20$\%$
of the peak performance on the SGI Altix 4700.
We are planning to implement better integration schemes and test other separations, for instance,
\begin{equation}
S_{\rm UV} = S_{g}, \hspace{3mm}
S_{\rm IR-1} = S_{f1}^{l}, \hspace{3mm}
S_{\rm IR-2} = S_{f2}^{l} + S_{fr}^{s} + S_{det}^{s} + S_{det}^{l}\,.
\end{equation}

\acknowledgments

~The numerical calculations have been performed on the SGI Altix 4700 at LRZ (Munich), 
as well as on the APEmille at DESY (Zeuthen). We thank all institutions. 
This work has been suppo-\\rted in part by the EU Integrated Infrastructure 
\hspace{-0.2mm}Initiative \hspace{-0.2mm}Hadron \hspace{-0.2mm}Physics (I3HP) under contract nu-\\mber \hspace{0.2mm}RII3-CT-2004-506078 and
by the \hspace{0.2mm}DFG under contract \hspace{0.2mm}FOR 465 
(Forschergruppe Gitter-Ha-\\dronen-Ph\"anomenologie).


\begin{thebibliography}{99}

\bibitem{JLQCDnf31st}
S. Aoki {\em et~al.},
\newblock Phys. Rev. {\bf D72} (2004) 054510 [hep-lat/0409016].

\bibitem{Symanzik}
K. Symanzik,
\newblock Nucl. Phys. {\bf B226} (1983) 187.

\bibitem{Morningstar}
C. Morningstar and M. J. Peardon,
\newblock Phys. Rev. {\bf D69} (2004) 054501 [hep-lat/0311018].

\bibitem{UVfilterDurr}
S. Capitani, S. D\"urr and C. Hoelbling, 
\newblock JHEP 0611 (2006) 028 [hep-lat/0607006].

\bibitem{Holger07100990}
H. Perlt {\it et al.} [QCDSF Collaboration],
PoS(LATTICE 2007)250 [arXiv:0710.0990].

\bibitem{hep-lat/0405026}
S. Boinepalli {\em et~al.},
\newblock Phys. Lett. {\bf B616} (2005) 196 [hep-lat/0405026].

\bibitem{hep-lat/0405015}
J.M.Zanotti {\em et~al.},
\newblock Phys. Rev. {\bf D71} (2005) 034510 [hep-lat/0405015].

\bibitem{Hasenbusch:2001ne}
M. Hasenbusch,
\newblock Phys. Lett. {\bf B519} (2001) 177 [hep-lat/0107019].

\bibitem{RHMC}
M. A. Clark and A. D. Kennedy,
\newblock Nucl. Phys. Proc. Suppl., {\bf 129} (2004) 850 [hep-lat/0309084].

\bibitem{AliKhan:2003br}
A. Ali Khan {\em et~al.},
\newblock Phys. Lett. {\bf B564} (2003) 235 [hep-lat/0303026].

\bibitem{Sexton:1992nu}
J. C. Sexton and D. H. Weingarten,
\newblock Nucl. Phys. {\bf B380} (1992) 665.

\bibitem{MICLfine}
C. Aubin {\em et~al.},
\newblock Phys. Rev. {\bf D70} (2004) 094505 [hep-lat/0402030].







\end{thebibliography}
\end{document}